# The ICARUS Front-end Preamplifier Working at Liquid Argon Temperature

B. Baibussinov [1], C. Carpanese [1], F. Casagrande [1,*], P. Cennini [2], S. Centro [1],
A. Curioni [3,†], G. Meng [1], C. Montanari [4], P. Picchi [5], F. Pietropaolo [1],
G. L. Raselli [4], C. Rubbia [2,6], F. Sergiampietri [7], S. Ventura [1].

[1] *Dipartimento di Fisica e INFN, Università di Padova, via Marzolo 8, Padova, Italy*
[2] *CERN, European Laboratory for Particle Physics, CH-1211 Geneva 23, Switzerland*
[3] *Dipartimento di Fisica e INFN, Università di Milano, via Celoria 16, Milano, Italy*
[4] *Dipartimento di Fisica e INFN, Università di Pavia, via Bassi 6, Pavia, Italy*
[5] *Laboratori Nazionali di Frascati (INFN), via Fermi 40, Frascati (RM), Italy*
[6] *Laboratori Nazionali del Gran Sasso (INFN), s.s.17bis km18.910, Assergi (AQ), Italy*
[7] *INFN Sezione di Pisa, via Livornese 1291, San Piero a Grado (PI), Italy*

ABSTRACT: We describe characteristics and performance of the low-noise front-end preamplifier used in the ICARUS 50-litre liquid Argon Time Projection Chamber installed in the CERN West Area Neutrino Facility during the 1997-98 neutrino runs. The preamplifiers were designed to work immersed in ultra-pure liquid Argon at a temperature of 87 K.

## 1 - Introduction

The successful operation of the ICARUS T600 detector at LNGS [1] and the recent growing interest in the potentiality of the LAr-TPC as multi-kton detector for neutrino physics and rare events search, have triggered additional R&D studies to further improve the detector performance with respect to the presently available technologies. In particular the use of low noise front-end electronics working immersed in liquid Argon is under discussion and test in several laboratories. In the present note, a description is given of the solution developed in the late nineties within the ICARUS collaboration, and successfully operated for several years in a 50-litre LAr-TPC. The pre-amplifier design and its performance are discussed in detail.

The 50-litre liquid Argon Time Projection Chamber (LAr-TPC) is a prototype, which was built in the framework of the R&D activity for the ICARUS experiment. It has been successfully operated in the CERN WANF neutrino beam in the years 1997 and 1998 [2]. In the following years it has been upgraded to test long drift distances (up to 1.5 meters) [3] and scintillation light as T=0 signal [4] and to optimise the new front-end electronics to be mounted on the first ICARUS T600 module [5].

In the ICARUS LAr-TPC, ionisation electrons produced in the active volume, drift toward the read-out anode, under a uniform electric field ranging up to 500 V/cm. The signal sensed by the read-out electrodes is the current induced by the electrons drifting near by (no signal multiplication occurs in liquid Argon).

---

[*] Present address: *Oak Ridge National Laboratory, Oak Ridge TN 37831-6469, USA*
[†] Present address: *Institute for Particle Physics, ETH Hönggerberg, CH-8093 Zürich, Switzerland*

In the case of the 50-litre LAr-TPC the active volume is delimited by the anode and the cathode surfaces, 325 by 325 mm$^2$ each, and by four vetronite walls, 475 mm in length, supporting the field shaping electrodes. The anode is made of two read-out planes spaced by 4 mm. Each plane is made of a grid of 128 inox wires (100 μm diameter, 2.54 mm pitch); the wires direction of one plane is orthogonal to that of the other plane. The electron drift velocity between cathode and anode and between anode planes is of the order of 1.5 mm/μs, corresponding to a collection time of 3-4 μs and a total drift time of ~300 μs.

The schematic layout of the ICARUS read-out electrodes is shown in Figure 1. The typical signals due to ionising tracks running perpendicular to the electric field direction and sensed by the wires of the two read-out planes are also shown.

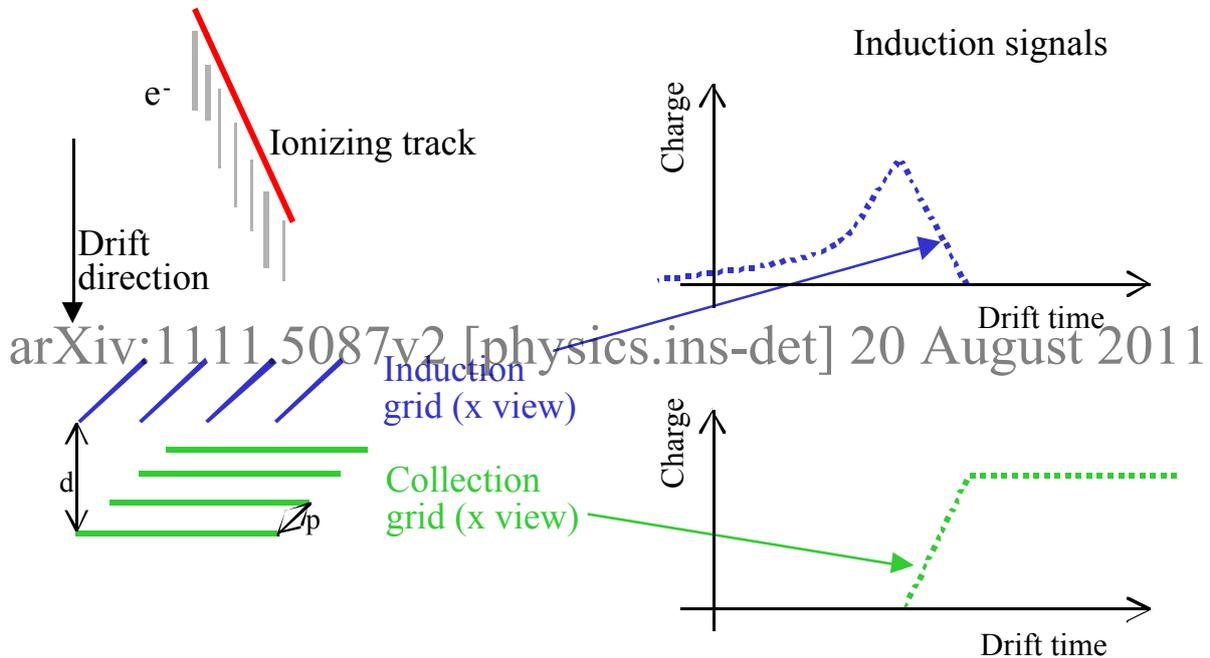

**Figure 1. - Left**: Schematic layout of the ICARUS read-out electrodes. In the case of the 50-litre LAr-TPC, d = 4 mm and p = 2.54 mm. **Right**: typical charge signals induced on the wires of the two read-out planes by ionising tracks running perpendicular to the electric field direction.

The charge induced on a wire by a minimum ionising particle (MIP) is 2.5 fC or ~15000 electrons. Each wire is read-out by a front-end analogue amplifier and then digitised by an 8-bit Flash ADC with sampling frequency of 2.5 MHz. The digitised waveform is continuously stored in a circular memory buffer whose depth is enough to contain at least one drift length (up to 2 kbyte per channel in the case of the 50-litre LAr-TPC).

Only when a trigger occurs and after a delay equivalent to one drift time, the content of the circular buffers is transferred to the mass storage device. This mode of operation makes the detector fully and continuously sensitive.

In order to exploit at best the excellent characteristics of liquid Argon as tracking and calorimetric medium, a large Signal-to-Noise level is needed (S/N > 10). Hence very low noise preamplifiers are required (ENC ~ 1000 electrons).

Since the electronic noise increases with increasing input capacitance, in the 50-litre LAr-TPC it was decided to mount the preamplifiers *immersed in liquid Argon* directly connected at the wire end after capacitive de-coupling to minimize the input capacitance.

The aim of the exposure of the 50-litre LAr-TPC at the CERN neutrino beam was to gain experience with real neutrino events and find the best configuration of the front-end analogue electronics. Following the results obtained in the off-line analysis the characteristics of the final configuration of the electronics for the ICARUS T600 module have been defined [6].

**2 -The preamplifier working at liquid Argon temperature.**

Independently from other considerations common to the design of low noise preamplifiers, the first choice concerned the technology that can operate at liquid Argon temperature, namely 87 K. Dedicated studies carried on in the past [7] [8] [9] showed that the choice can be limited to three construction processes available today:

- Silicon Junction Field Effect Transistors: S-J-Fet;
- Gallium Arsenide Field Effect Transistors: GAs-Fet;
- Metal Oxide Silicon Field Effect Transistors: MOS-Fet.

In the frequency bandwidth required for the ICARUS detector (÷MHz), the best performances in terms of 1/f noise are obtained with the Silicon J-Fet.

The high Gate impedance of a J-Fet and its very low leakage current made this device ideal as first stage of the Preamplifier. In the following stages it is not evident that J-Fet's are the most suited components, nevertheless the capability of this technology to work at cryogenic temperatures makes the circuit design, based on J-Fet's in all stages, appealing.

The widely used Silicon J-Fet's show their best characteristics at around 150 K; at lower temperatures their characteristics are drastically worsened. However today High Resistance Substrate devices, designed to work reliably at 87 K [10] [11], are available on the market.

The reliability of electronic components at cryogenic temperature and the heat introduced in the Detector's volume are very important issues to be considered in the circuit design. Extensive tests made at BNL and NASA [12] demonstrate that if some standard rules are followed, like the choice of the right circuit support, the extensive burn-in and the temperature cycling before the final installation, the risks of malfunctioning are very low. Concerning the boiling-off generated by the heat produced by the preamplifier, a test made for LAr Calorimeters [13] on circuits with a surface of ~1.5 cm$^2$ and a power dissipation of 70 mW, gave clear indications that there are no boiling effects for power dissipation lower than 100 mW/cm$^2$.

The tests we have made in the laboratory, the experience acquired on small prototypes [14] and the results obtained by other groups [15], demonstrate that the use of the front-end Electronics immersed in Liquid Argon is possible and reliable.

## 2.1 - Noise Theoretical Aspects

In a first attempt the basic scheme retained for the first stage of the front-end Electronics is that of an Integrator. The Equivalent Noise Charge, ENC, referred to the preamplifier input, depends on the following parameters, as described in [16]:
- Input capacitance;
- Trans-conductance of the first preamplifier stage;
- Temperature;
- Equivalent Noise Bandwidth.

By definition, the thermal noise is a function of temperature and it decreases with the temperature as far as passive components are concerned.

In the following, all numbers will be given assuming a bandwidth dominated by a transfer function of the type $x^2e^{-x}$ (where $x = t/T$ and $T = RC = 1.5$ μs) that has a maximum at $x = 2$ ($T = 3$ μs). This function corresponds to a filter composed of a CR differentiator followed by two RC integrators. Although this kind of filtering will not be applied in the LAr-TPC, this shaping is useful to compare the performances of the proposed circuit within a defined bandwidth.

The total ENC, expressed in number of electrons, is function of two terms: $ENC_p$ the Parallel noise and $ENC_s$ the Serial Noise; both are temperature dependent while only the Serial Noise depends, linearly, on the Input Capacitance. The Input capacitance $C_{in}$ is the sum of the Detector capacitance, Input Cables capacitance and Preamplifier internal capacitance. In case the input device of the Integrator is a J-Fet, the formula that applies is:

$$ENC^2 = (2kT\,P/R_p)^2 + (2kT\,R_s\,S)^2$$

where:
- k is the Boltzmann's constant;
- T the temperature in K;
- P and S are the coefficients defined by the filter;
- $R_p$ is the equivalent Input parallel resistance;
- $R_s$ is the equivalent Input serial resistance ($0.7/g_m$) [17].

For detectors with a large input capacitance, the serial noise dominates. As a consequence a good approximation for the noise formula is:

$$ENC = \sim ENC_0 + N_0\,C_{in}$$

where:
- $ENC_0$ is the noise for $C_{in} = 0$ expressed in number of electrons;
- $N_0$ is the Noise Slope expressed in number of electrons per pF.

Since in our application the detector capacitance is very small (< 50 pF), the Noise Slope is not an important parameter. Therefore the use of J-Fet's with a moderate trans-conductance $g_m$ is acceptable.

## 2.2 - The TOTEM Structure

The guidelines used for the circuit design are all going in the direction of minimizing the following parameters: number of components, noise, and power consumption. The circuit is an integrator with a feedback capacitor of 2.2 pF and feedback resistor of 100 MΩ. The sensitivity at $V_{out}$ is 0.45 mV/fC and the decay time-

— 4 —

constant 220 µs. In Figure 2 three possible structures envisaged for the preamplifier are reported.

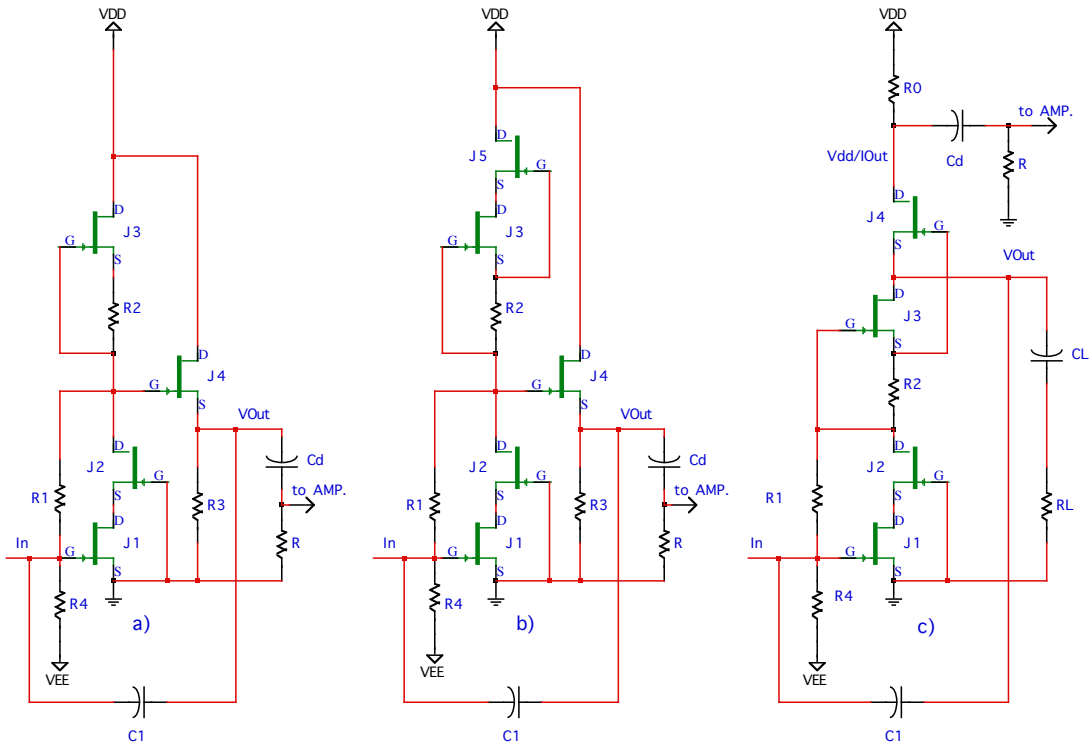

**Figure 2:** Three possible structures envisaged for the ICARUS preamplifier (a) basic configuration, (b) basic with cascode current generator, (c) TOTEM.

Version (a) consists of a current source, made with J3-R2, which generates a current flowing in the input stage (J1-J2) configured as an unfolded Cascode. The current is set to a value requiring a $V_{gs} \geq 0.7$ V. Since the four J-Fet's are all of the same type, the J1 drain is polarized with a voltage equal to $V_{gs}$. The geometry of the NJ132 process from InterFet, allows the correct working outside the saturation region for currents up to 5 mA. The polarization of J2 drain, determined by R1, R2 and $V_{ee}$, is set to approximately half of the $V_{dd}$ value. The feedback capacitance, C1, is connected on the output buffer made with the source-follower J4-R3. The DC polarization on the $V_{out}$ terminal requires a capacitive coupling to the following stage. Unfortunately the poor performance of the output buffer does not allow driving a transmission line.

The layout of configuration (b) is an improved version of (a) in which the current generator is made with a Cascode (J3-J5-R2). The benefit of this scheme concerns essentially the open-loop gain that depends on the equivalent impedance seen by the J2 drain.

The above two versions present the following disadvantages: *Inadequate Driving Performance and large Power Consumption*. A possible improvement consists in setting the current in J4 equal to that flowing in J1-J2, so that the output stage requires only half of the power needed by the total circuit. Unfortunately the additional stage, able to drive in differential mode a transmission line, needs most likely more components and more power than the Preamplifier itself. Therefore, the solution to add an extra output stage has been discarded.

Circuit (c), called TOTEM, shows all the characteristics needed by a circuit working at LAr temperature:
• Minimization of active components;
• Ability to drive a transmission line;
• Reduced Power Consumption.

The circuit, optimised for low capacitance detectors, requires only four J-Fet's. Instead of using the differential voltage-output, TOTEM has a current-output. The advantages of the current-driven transmission lines are not to be demonstrated.

The TOTEM architecture is very similar to that previously discussed for scheme (b), the difference is on the point at which the voltage output is taken.

If nothing is connected on the $V_{out}$ terminal, the current flowing through the Power-Supply is equal to the sum of the current defined by the Cascode current-generator J4-J3-R2, plus the current flowing in R1. Since the DC polarization current is typically 1.6 mA, the current in R1 and the detector current in the nA region, it can be assumed that the output current is only defined by the internal current generator. Consequently the current on the $V_{dd}$ terminal is constant independently from the input signal.

If a resistive load is connected to the output, the current in the $V_{dd}$ terminal has the shape of the voltage at $V_{out}$ and a value equal to $V_{out}/R_{load}$.

To avoid unwanted power dissipation $R_{load}$ is capacitively coupled to ground. The additional advantage of this configuration is the power supply distribution made individually to every circuit through the output connection. Due to the very large impedance (100 MΩ) on $V_{ee}$ and the de-coupling on $V_{ee}$, the Preamplifiers are practically all connected together only through a single common point: the ground. This situation is very safe concerning the risks of oscillations of a large distributed system.

*2.3 - The Real Circuit*

The circuit performance depends on the J-Fet characteristics and therefore the choice of the suitable components is particularly delicate. For the application in the ICARUS detector, the two most important parameters are the ratio Gate Capacitance versus Trans-conductance $C_{gs}/g_m$ and Trans-conductance versus Drain Current $g_m/I_d$.

The circuit has been designed assuming the use of the J-Fet IF1330 from INTERFET, made with a process called High Resistive Substrate that ensures the correct functioning at LAr temperature. The most important characteristics of this device are: $g_m$ = 15 mA/V at $I_{dss}$; $C_{gs}$ = 20 pF; $g_m/I_d$ = ~ 4. Figure 3 shows the circuit schematics as implemented on a Hybrid with a surface of about 4 cm$^2$.

The polarization current defined by J3-R3 is typically 1.6mA and gives a $g_m$ of 10mA/V. The power consumption of the preamplifier depends on the dynamic range required, according to the formula: $P = I_0 * (4 + V_{dvn})$ where $I_0$ is the polarization current and $V_{dvn}$ the output dynamic range. Assuming a minimum signal of 1.13 mV (2.5 fC on 2.2 pF), the power consumption for a dynamic range of ~1300 ($V_{dvn}$ = ± 1.5 V) is ~ 11 mW for a voltage of 7 V at the $V_{dd}/I_{out}$ terminal. According to [13] this heat production is far from producing boiling off of liquid Argon.

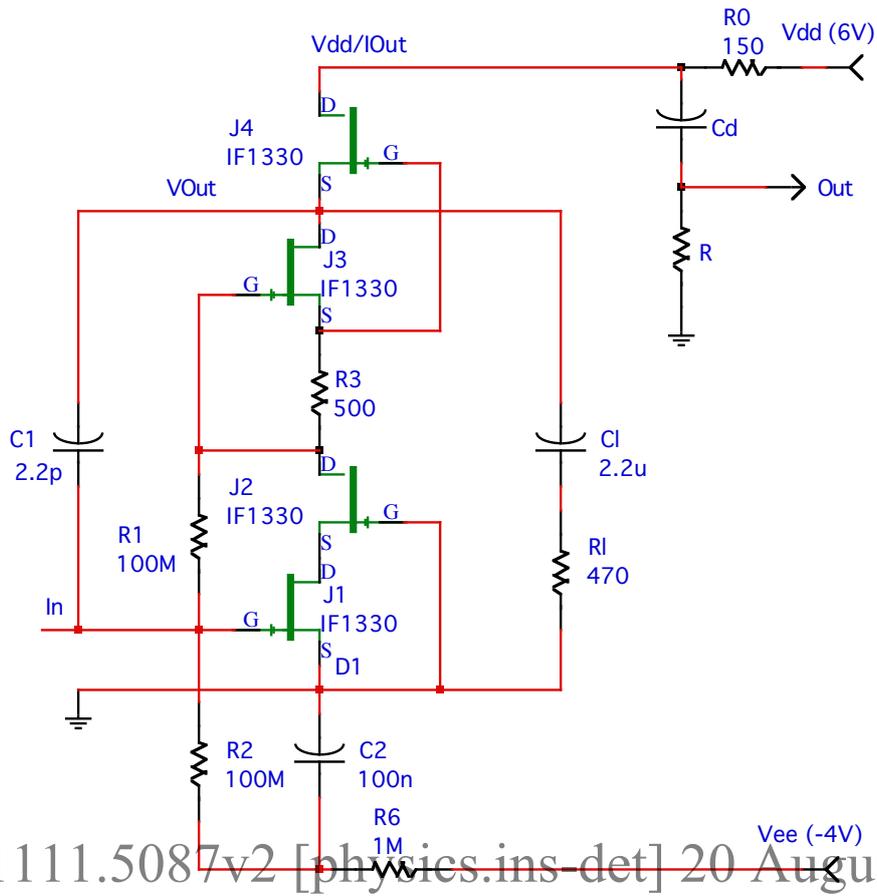

**Figure 3 -** The actual TOTEM Circuit.

In Figure 4, 5 and 6 the results of SPICE simulations of the TOTEM circuit are presented concerning some relevant parameters: the output frequency response, the input impedance and the wave form of the output current in response of a input charge of 10 fC and 3 μs rise time.

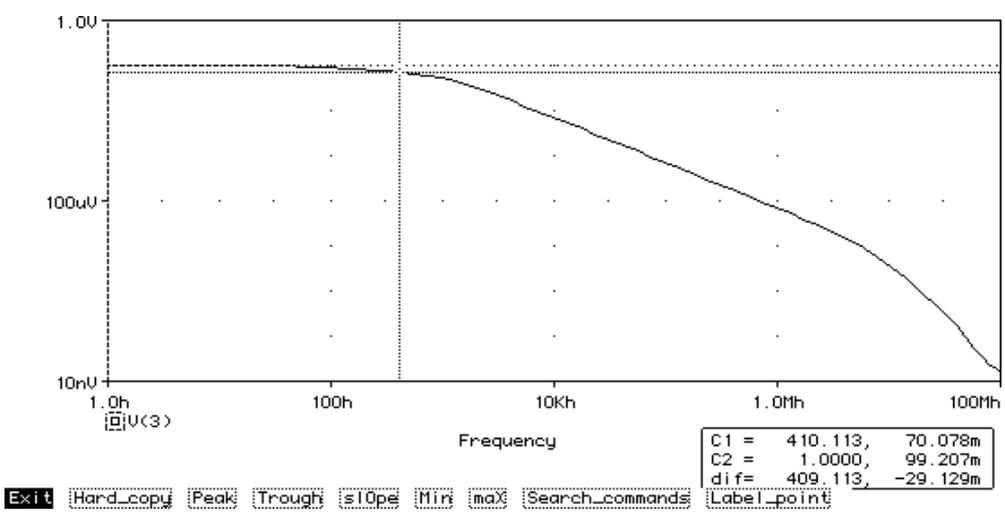

**Figure 4 -** Spice simulation of the Preamplifier frequency response.

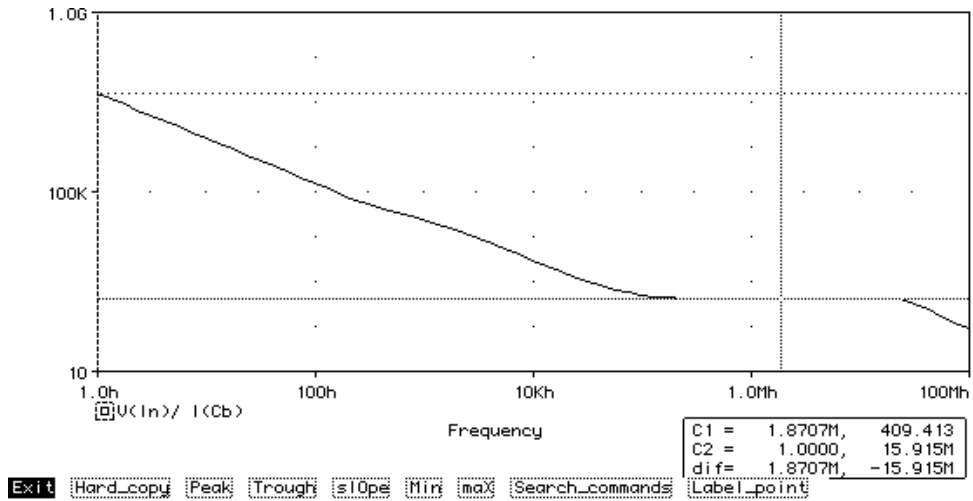

**Figure 5 -** Spice simulation of the Preamplifier Input impedance.

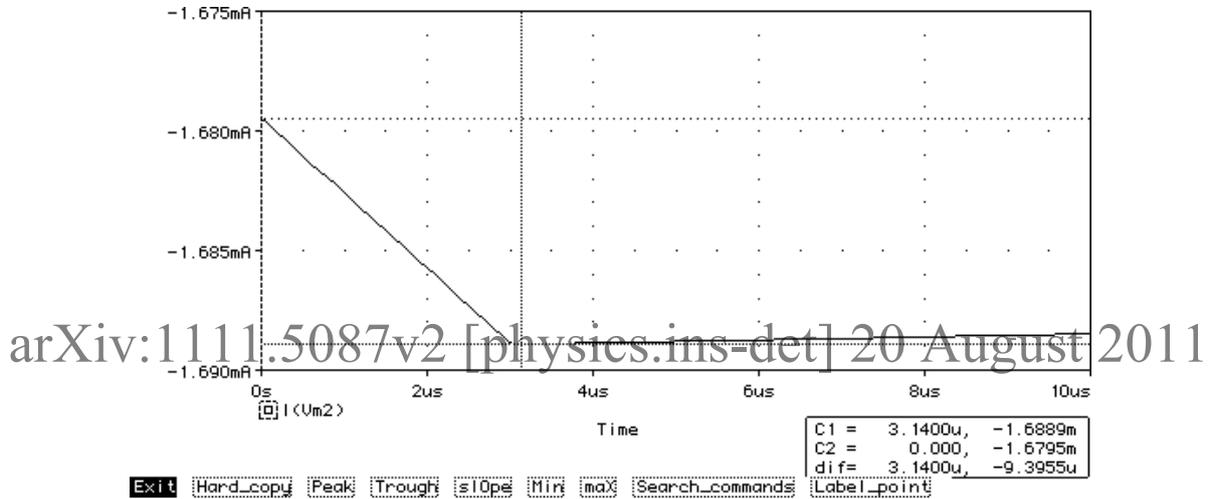

**Figure 6 -** Spice simulation of the Preamplifier Current output.

*2.4 - Circuit performances*

Before using the TOTEM preamplifier on the 50-litre LAr-TPC, extensive noise measurements have been performed using a complete read-out chain in which the TOTEM current output is injected through a time constant of 1.6 ms on the Amplifier circuit reported in Figure 7. The complete analogue chain delivers a signal to the Flash ADC's with a minimum rise-time of ~ 500 ns essentially determined by the Amplifier bandwidth of ~ 600 kHz. This limitation is chosen to satisfy the requirements of minimum noise and the maximum bandwidth for the triangular signals ($\geq 3$ μs rise and decay-time) issued by the read-out planes of the ICARUS LAr-TPC.

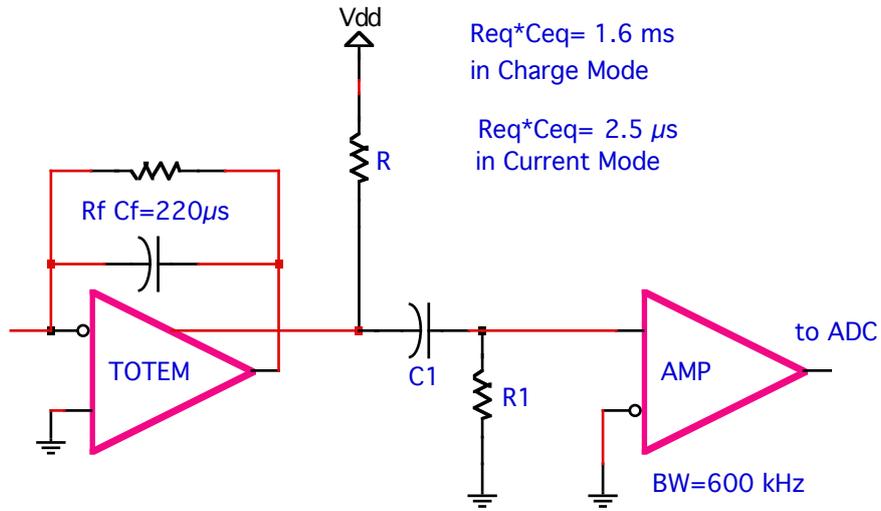

**Figure 7** - Block diagram for Charge and Current Digitisation.

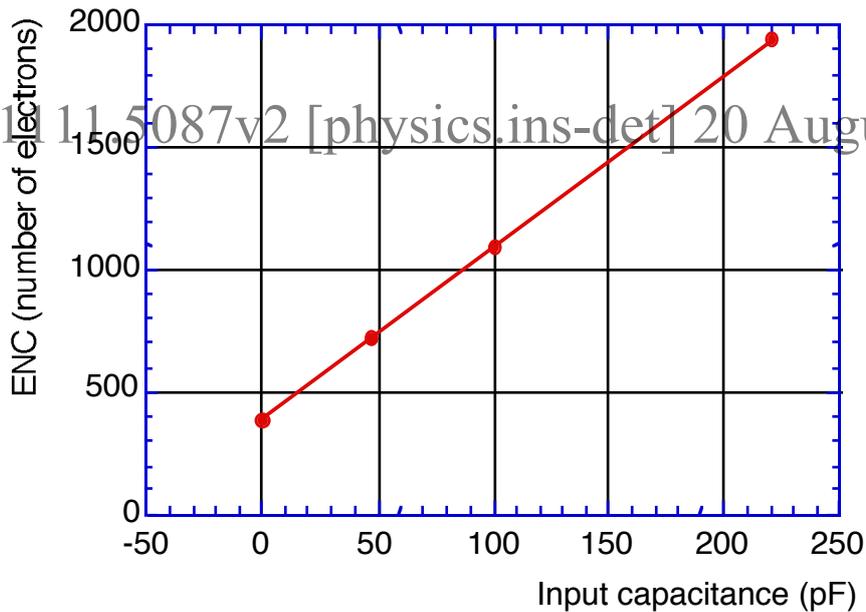

**Figure 8** – Measured Equivalent Noise Charge for the pre-amplifier working in charge mode at room temperature. The ENC at $C_{in}$ = 0 is 390 electrons and the ENC slope is 7 electrons/pF.

The ENC measurements were made injecting a charge of 2.5 fC with a rise time of 2.5 µs. The ADC conversion was made at a sampling time of 400 ns, corresponding to the normal ICARUS DAQ conditions. The pedestal of the digitised waveform was evaluated over a window of 40 samples and the ENC was calculated as the RMS of the maximum after pedestal subtraction. In Figure 8 the measurements are presented as a function of the detector capacitance.

The measured performances reported in Table 1, are given for the TOTEM Integrator working at room temperature. In LAr the ENC does not change significantly while the open-loop gain is enhanced and the input impedance reduced.

Table 1: Performance of the TOTEM integrator at room temperature.
(*) With a CR-RC$^2$ shaper, RC = 1.5 µs.

| | |
|---|---|
| Power Consumption | 11 mW |
| Dynamic Range | ±2.8 pC (or ±1.5 pC) |
| Linearity | ±0.5 % at ±0.5 V |
| Open Loop Gain | 150 |
| Input Impedance | 420 Ω |
| Sensitivity ($V_{out}$) | 0.45 mV/fC |
| Sensitivity ($I_{out}$) | 0.9 µA/fC |
| ENC ($C_d$=0) (*) | 390 electrons |
| ENC Slope (*) | 7 electrons/pF |
| Input Capacitance | 20 pF |
| Decay Time | 220 µs |

**4. - Charge Digitization versus Current Digitization.**

The problem of the dynamic range is intrinsically considered and solved in the original TOTEM design [18]. The solution consists in changing the load impedance on the Integrator voltage output ($C_{load}$ and $R_{load}$ in Figure 2). Ideally, if the load is a pure capacitance, the output current corresponds to the current necessary to charge $C_{load}$ at the voltage across $C_1$. The circuit behaves as a trans-impedance amplifier with a gain, current-out versus current-in, equal to the ratio $C_{load}/C_1$. The series resistance $R_{load}$, necessary for stability problems, introduces a small integration on the current shape.

In a circuit with $C_1$ = 2.2 pf, $C_{load}$ = 2.2 nF and $R_{load}$ = 500 Ω, the gain is 1000 and the band width 160 KHz.

The advantages of digitizing the current induced by the detector instead of the charge have been studied in detail. The energy reconstruction on the signals issued by the collection plane is made performing a digital sum of the digitized current.

The points in favour of this technique are:

- Large dynamic range on the reconstructed charge, avoiding saturation event in case of a simple 8 bit resolution Flash ADC.
- Good tracking and track separation.
- Suppression of the shadow effect allowing to distinguish and precisely measure a track issued by a MIP as part of an electromagnetic shower.
- Possibility to measure long tracks perpendicular to the wire planes.
- Suppression of the problems arising in case of low frequency microphonic noise.
- Noise performances similar to those achieved in charge-mode.

For the neutrino runs in 1997 and 1998 with the 50-litre LAr-TPC, it was decided to implement the current read-out just making a small change in the receiver electronics mounted outside the LAr-TPC reducing the coupling time constant of the Amplifier to 2.5 µs ($R_1$, $C_1$ in Figure 6).

In Figure 10 the Spice simulations under these conditions are reported. The current pulse is slightly integrated compared to the real waveform but the off-line analysis made on real tracks demonstrates that the bandwidth limitation is not affecting the energy resolution while keeping a good track separation, even in case of large showers.

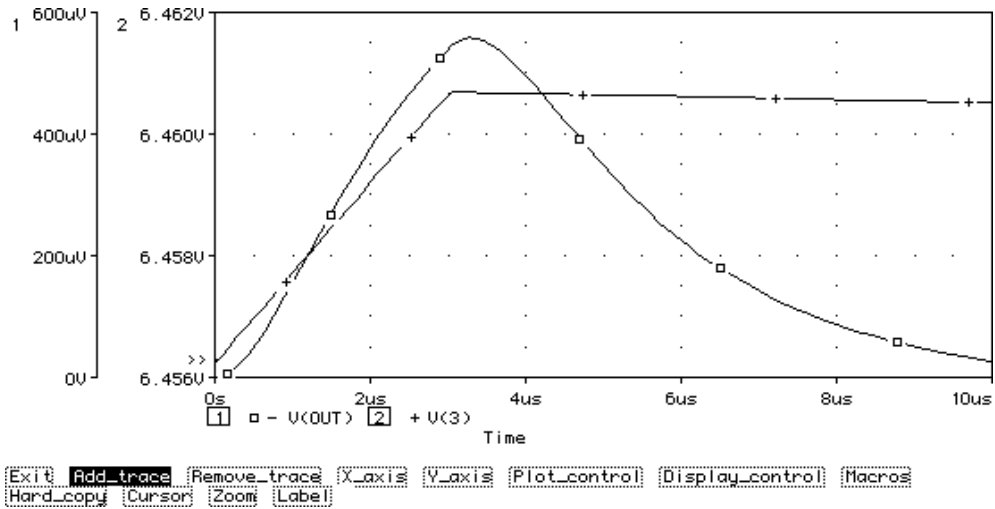

**Figure 10** - Spice simulation of the Preamplifier Voltage output & Amplifier output in current mode.

In Figure 11 the current signal issued by a charge of 2.5 fC at 3 µs rise time is shown.

The ENC measurements reported in Figure 12 are calculated in the following way: the base line is evaluated taking the mean over the 40 ADC samples preceding the pulse, than the following 30 samplings are added together after the base line subtraction.

The noise behaviour in Current mode is practically the same to that measured in Charge mode. A measuring campaign was made to set the specifications for the new front-end electronics working at room temperature for the large ICARUS modules. The results have been reported in [2]. Here we simply report that even using a Trans-impedance Preamplifier followed by an amplifier with a bandwidth limitation at 40dB/decade, there is no difference between the noise measured digitizing the current and that expected in Charge mode.

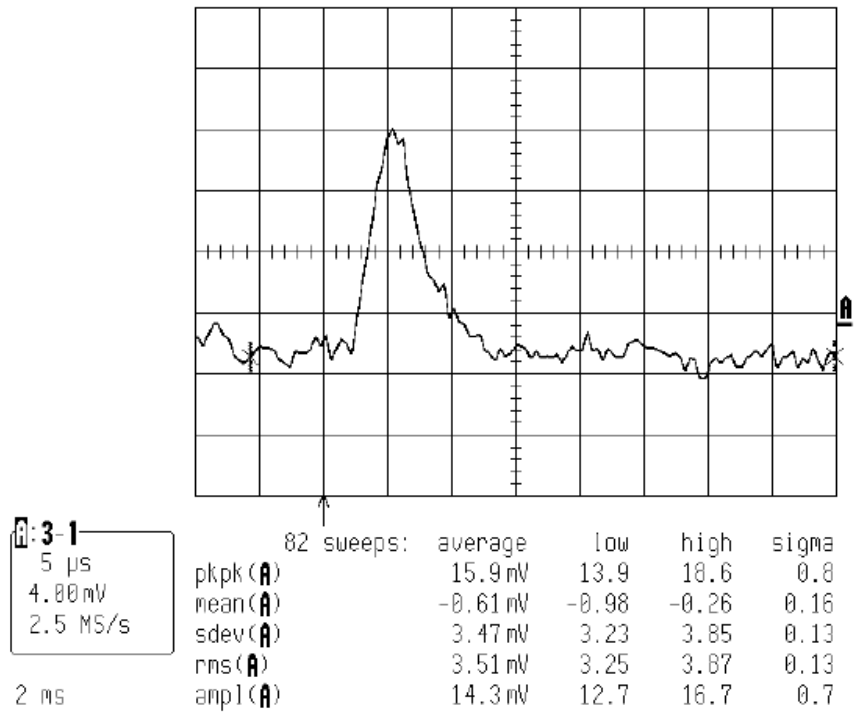

**Figure 11** - Signal in current mode at the ADC input for $Q_{in}$=2.5 fC and $t_r$= 3 µs (peak at 12 ADC counts).

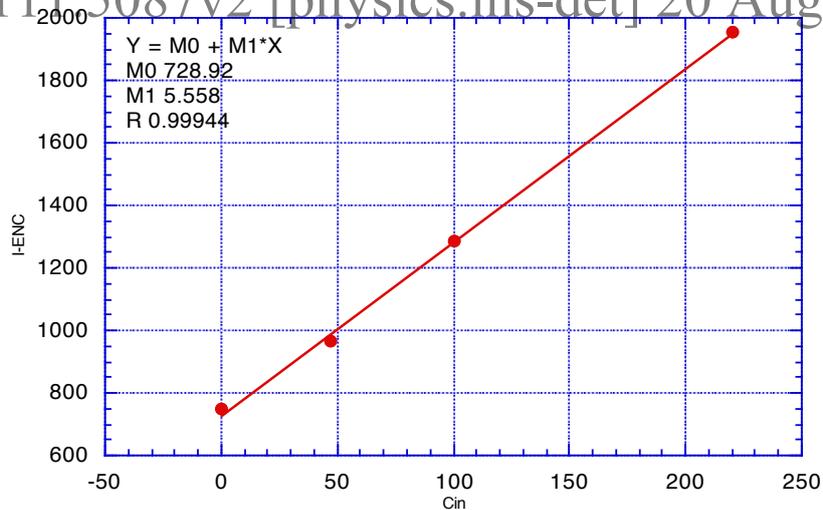

**Figure 12**: Equivalent Noise Charge in Current mode (BW = 600 kHz) after digital Integration made on 30 samples at 400 ns sampling rate.

*4.1 Experience gained with the front-end electronics in the 50-litre LAr-TPC*

As previously mentioned, for the CERN WANF neutrino runs in 1997 and 1998 the 50-litre LAr-TPC was equipped with 256 TOTEM pre-amplifiers directly connected to the read-out wires and immersed in ultra-pure liquid Argon. We collected several hundred thousand beam-related events, including neutrino interactions and through-going muon tracks. In Figure 13 we show, as an example, the collection view 2D image of a high multiplicity neutrino charged current interaction.

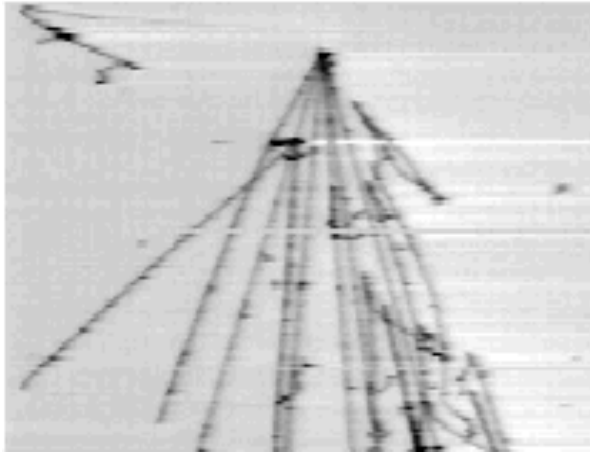

**Figure 13** - An example of recorded neutrino interaction in the 50-litre LAr-TPC located at the CERN neutrino beam in 1997. The neutrino comes from the top of the picture. The horizontal axis is the time axis (drift direction) and vertically is the wire number. The visible area corresponds to 475 x 325 mm$^2$.

We recall that the integration time of the read-out chain was set to 2.5 μs. It was chosen smaller than the typical signal rise-time, 3-4 μs, in order not to distort the waveform shape, and large enough to get the best possible signal-to-noise ratio (S/N ~ 11 with a signal FWHM ~ 4 μs). This allowed setting the MIP signal at 12 ADC counts.

The choice was satisfactory also because the risk of pile-up in events containing electromagnetic showers was highly suppressed. This essential feature is due to the fact that the duration of each MIP signal was comparable with the distance between tracks in electromagnetic showers. A further consequence of the front-end choice was that the digital dynamic range of 8 bits was sufficient even for the case of electromagnetic showers.

From the inspection of many similar events and the comparison with detector simulations we extracted and carefully studied fundamental properties of the LAr-TPC as ionization detector such as calorimetric energy measurement, particle identification through dE/dx vs. range, single point space resolution, track separation. Full reconstruction of neutrino interactions vertexes was also possible leading to high quality measurements of quasi-elastic muon neutrino events kinematics and cross-sections [19].